\documentclass{pasj01}

\begin{document} 
\Received{}
\Accepted{}

\title{Turbulent Cosmic-Ray Reacceleration and the Curved Radio Spectrum
of the Radio Relic in the Sausage Cluster}

\author{Yutaka \textsc{Fujita}\altaffilmark{1}}
\altaffiltext{1}{Department of Earth and Space Science, 
Graduate School of
 Science, Osaka University, Toyonaka, Osaka 560-0043, Japan}
\email{fujita@vega.ess.sci.osaka-u.ac.jp}

\author{Hiroki \textsc{Akamatsu},\altaffilmark{2}}
\altaffiltext{2}{SRON Netherlands Institute
for Space Research, Sorbonnelaan 2, 3584 CA Utrecht, The Netherlands}

\author{Shigeo S. \textsc{Kimura}\altaffilmark{3}}
\altaffiltext{3}{Frontier Research Institute for Interdisciplinary 
Sciences, 
Tohoku University, Sendai 980-8578, Japan
and Astronomical Institute, Tohoku University, Sendai 980-8578, Japan}


\maketitle

\begin{abstract}
It has often been thought that the northern radio relic in the galaxy
cluster  CIZA~J2242.8$+$5301 (the ``Sausage'' Cluster) is associated with
cosmic ray (CR) electrons that are accelerated at a shock through the
diffusive shock acceleration (DSA) mechanism. However, recent radio
observations have shown that the radio spectrum is curved, which is
inconsistent with the prediction of a simple DSA model. Moreover, the CR
electron spectrum before being affected by radiative cooling seems to be
too hard for DSA. In this study, we show that these facts are natural
consequences if the electrons are reaccelerated in turbulence downstream
of the shock. In this model, DSA is not the main mechanism for
generating high-energy electrons. We find that the mean free path of the
electrons should be much shorter than the Coulomb mean free path for
efficient reacceleration. The scale of the turbulent eddies must be
smaller than the width of the relic. We also predict hard X-ray spectra
of inverse Compton scattering of photons.
\end{abstract}

\section{Introduction}

Radio relics are synchrotron emissions often found on the periphery of
clusters of galaxies. From their positions and prolonged shapes, they
have been associated with cosmic ray (CR) electrons accelerated at
shocks in the intracluster medium (ICM). In fact, shocks have been
discovered at some relics by X-ray observations
\citep{fin10,mac11a,aka12a,aka12b,aka13a,ogr13b,bou13a}.  It has often
been believed that the CR electrons are accelerated at the shocks
through the diffusive shock acceleration (DSA) mechanism, which is
first-order Fermi acceleration
(e.g. \cite{roe99a,fuj01b,kan12a,yam15a}).  However, recent observations
have shown that the acceleration is much more complicated than
expected. For example, the radio spectrum of the relic in the galaxy
cluster 1RXS J0603.3 $+$4214 (``Toothbrush'' cluster) is hard and
curved, which contradicts predictions based on a simple DSA model
\citep{van12b,str16a}.  Moreover, XMM-Newton observations showed that
the position of the shock does not coincide with that of the relic
\citep{ogr13a}, although the shock-relic offset was not recognized in
deeper Chandra observations \citep{van16a}.

In \citet{fuj15b}, we showed that the discrepancies are solved if
electrons are reaccelerated in turbulence {\it downstream} of the
shock. In this model, the electrons are first accelerated at a shock
through the DSA, although the efficiency does not have to be high. Then,
the seed CR electrons are efficiently reaccelerated in the turbulence
through second-order Fermi acceleration. Turbulent reacceleration has
been studied to explain the origin of giant radio halos observed in
clusters (e.g. \cite{sch87a,bru01a,pet01b,fuj03a}). \citet{fuj15b}
showed that it can also be applied to radio relics. Turbulent
(re)acceleration behind shocks has been considered for objects such as
supernova remnants (e.g. \cite{ino10a,zha15a}) and the Fermi bubbles
(e.g. \cite{mer11a,sas15a}).

While the relic in the Toothbrush Cluster has a complicated shape and
the Mach number of the associated shock is small (${\cal M}\lesssim 2$;
\cite{ita15a}), the northern relic in the cluster CIZA~J2242.8$+$5301
(the ``Sausage'' Cluster; $z=0.1921$) has a clear bow-like shape
\citep{van10a} and the Mach number is relatively large (${\cal M}\sim
3$; \cite{aka15a}). This may indicate that the DSA model is preferable.
However, even this ``textbook'' example of a relic has a hard, curved
radio spectrum, which is not easy to reconcile with DSA models
\citep{str16a}. We note that there are modified DSA models in which
fossil electrons upstream of the shock are reaccelerated at the shock
through DSA (e.g. \cite{kan15b}) or magnetic fields vary
\citep{don16}. These could result in a curved radio spectrum. In
addition, the impact of the Sunyaev-Zel’dovich effect on the radio
spectrum could not be ignored \citep{bas15a}. However, in this study, we
attempt to apply our reacceleration model to the relic in the Sausage
Cluster in order to solve the contradictions. We do not go into the
details of the micro-physics of turbulent reacceleration (e.g., the
cascade of turbulence, plasma instabilities, and so on), which are
highly unknown at present. Instead, we focus on finding out the
parameters that are consistent with observations.  The results will be
helpful when more comprehensive models are constructed in the future.

\section{Models}

In the model of \citet{fuj15b}, electrons are first accelerated at a
shock through DSA. The momentum spectrum of the accelerated electrons is
\begin{equation}
\label{eq:DSA}
 n_0(p) = A_0 p^{-s}\:,
\end{equation}
where $p$ is the electron momentum and $A_0$ is the normalization. The
index is given by
\begin{equation}
 s=\frac{r_c+2}{r_c-1}\:,
\end{equation}
where $r_c$ is the compression ratio:
\begin{equation}
 r_c=\frac{(\gamma_g+1){\cal M}^2}{(\gamma_g-1){\cal M}^2 + 2}\:,
\end{equation}
where $\gamma_g (=5/3)$ is the adiabatic index, and ${\cal M}$ is the
Mach number of the shock (e.g. \cite{bla87a}). The downstream density
$\rho_{\rm d}$ and the velocity $V_{\rm d}$ are related to the upstream
density $\rho_{\rm u}$ and the velocity $V_{\rm u}$ as $\rho_{\rm d}=r_c
\rho_{\rm u}$ and $V_{\rm d} = V_{\rm u}/r_c$, respectively. We assume
that the minimum momentum of the accelerated electrons is $p_{\rm
min}=m_e c$, where $m_e$ is the electron mass, and $c$ is the light
velocity. The normalization $A_0$ in equation~(\ref{eq:DSA}) is given as
follows. The kinetic energy flux upstream of the shock is $\rho_{\rm u}
V_{\rm u}^2/2$. We assume that a fraction $\chi_e$ of the flux is
consumed to accelerate electrons. This means that the kinetic energy
density of the accelerated CR electrons just downside of the shock is
$\epsilon_{e,sh}=\chi_e\rho_{\rm u} V_{\rm u}^3/(2 V_{\rm d}) =
\chi_e\rho_{\rm u} V_{\rm u}^2 r_c/2$. On the other hand, the density is
also written as
\begin{equation}
  \epsilon_{e,sh}=\int_{p_{\rm min}}^\infty m_e c^2 (\gamma-1) n_0(p) dp
\:,
\end{equation}
where $\gamma$ is the Lorentz factor of a CR electron. From these
equations, we obtain
\begin{equation}
 A_0 = \frac{\chi_e\rho_{\rm u} V_{\rm u}^2 r_c}{2 m_e c^2}
/\int_{p_{\rm min}}^\infty (\gamma-1) p^{-s} dp \:.
\end{equation}

We assume that turbulence develops downstream of the shock. The CR
electrons accelerated at the shock are swept downstream with the ICM and
they are reaccelerated in the turbulence. The evolution of the momentum
spectrum $n(t,p)$ on the comoving frame is calculated using the
Fokker-Plank equation,
\begin{equation}
\label{eq:evo}
 \frac{\partial n}{\partial t}
-\frac{\partial}{\partial p}\left(p^2 D_{pp}
\frac{\partial}{\partial p}\frac{n}{p^2}\right)
+ \frac{\partial}{\partial p}\left(\frac{dp}{dt}n\right) = 0\:,
\end{equation}
where $D_{pp}$ is the diffusion coefficient for momentum. The initial
spectrum is given by $n(t=0,p)=n_0(p)$. Since the CR electrons move with
the ICM, the distance of the electrons from the shock is written as
$x=V_{\rm d} t$.

We assume that the electrons are scattered by Alfv\'en waves because
they can develop even on small scales through the cascade of turbulence
and/or plasma instabilities (see Section~2.2.2 of
\cite{bru14a}). Thus, the diffusion coefficient for momentum is
\begin{equation}
\label{eq:Dpp}
 D_{pp}\sim \frac{1}{9}p^2 \frac{v_{\rm A}^2}{D_{xx}}\:,
\end{equation}
where $v_{\rm A}$ is the Alfv\'en velocity, and $D_{xx}$ is the spatial
diffusion coefficient,
\begin{equation}
\label{eq:Dxx}
 D_{xx}\sim \frac{c l_{\rm mfp}}{3}\:,
\end{equation}
where $l_{\rm mfp}$ is the mean free path of the electrons
(\cite{ohn02a}; see also \cite{ise87a,sch89a,fuj03a,bru04a}). The
Alfv\'en velocity is written as $v_{\rm A}=B_{\rm d}/\sqrt{4\pi\rho_{\rm
d}}$, where $B_{\rm d}$ is the downstream magnetic field.  When
particles are scattered by magnetosonic waves, $l_{\rm mfp}$ may be
comparable to the Coulomb mean free path of thermal particles $l_{\rm
mfp,C}$ (e.g. \cite{zuh13a}; see also \cite{bru07a}). However, when they
are scattered by Alfv\'en waves, $l_{\rm mfp}$ may be much smaller than
$l_{\rm mfp,C}$. Thus, we assume that $l_{\rm mfp}$ has the form of
\begin{equation}
\label{eq:lmfp}
 l_{\rm mfp}(t,p)=\eta(t) (p/p_0)^{2-q} l_{\rm mfp,C}\:,
\end{equation}
where $\eta(t)$ is the reduction factor, $p_0$ is the reference
momentum, and $q$ is the parameter representing the property of the
turbulence. The Coulomb mean free path $l_{\rm mfp,C}$ depends on the
downstream density $\rho_{\rm d}$ and temperature $T_{\rm d}$. Following
\citet{fuj15b}, we set $p_0=10^4 m_e c$ and $q=5/3$ (Kolmogorov case).
It is to be noted that even if we adopt $q=2$ (hard sphere
approximation), the results do not change qualitatively. We expect that
the turbulence and the Alfv\'en waves decay with the distance from the
shock and we include that effect in $\eta(t)$. Assuming that the decay
scale in space is $L_{\rm t}$, its time scale is $t_0=L_{\rm t}/V_{\rm
d}$. Thus, the reduction factor is represented by
\begin{equation}
\label{eq:eta}
 \eta(t)  = \eta_{\rm min} \exp(t/t_0)\:.
\end{equation}
This means that the mean free path increases as CR electrons move away
from the shock with the ICM. In equation~(\ref{eq:evo}), $dp/dt$
represents cooling of the CR electrons.  We include synchrotron, inverse
Compton scattering of cosmic microwave background photons, non-thermal
bremsstrahlung, and Coulomb interaction
\citep{gou75a,stu97a,sar99a,yan09}. The synchrotron emission depends on
$B_{\rm d}$.

\section{Results}
\label{sec:result}

X-ray observations have shown that the shock parameters for the northern
relic in the Sausage Cluster are $V_{\rm u}=2300\rm\: km\: s^{-1}$,
${\cal M}=2.7$, $\rho_{\rm d}=9.4\times 10^{-28}\rm\: cm^{-3}$, and
$T_{\rm d}=8.5$~keV \citep{aka15a}. The Coulomb mean free path
downstream of the shock is $l_{\rm mfp,C}\sim 45$~kpc. Since the spatial
scale of turbulence should be much smaller than the width of the relic
($\sim 100$~kpc), we assume that $L_{\rm t}=10$~kpc ($t_0=12$~Myr). We
choose the reduction factor at the shock ($\eta_{\rm min}=1.6\times
10^{-7}$), the acceleration efficiency ($\chi_e=5.0\times 10^{-7}$), and
the downstream magnetic field ($B_{\rm d}=0.30\:\mu\rm G$) so that the
resultant radio profile and spectrum are consistent with observations
(see later). In general, a smaller $\eta_{\rm min}$ and/or a larger
$B_{\rm d}$ give a larger $D_{pp}$, which leads to more efficient
reacceleration [equations~(\ref{eq:Dpp})--(\ref{eq:eta})].

\begin{figure}
 \begin{center}
  \includegraphics[width=8cm]{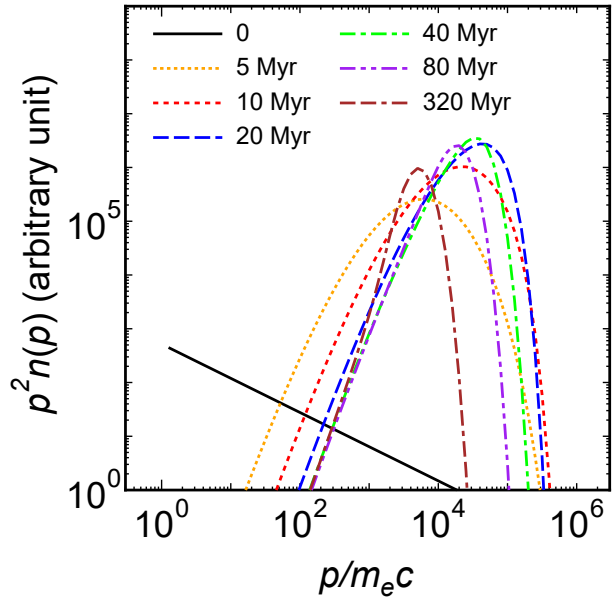} 
 \end{center}
 \caption{Electron spectra behind the
shock. Time $t$ is shown in the legends; $t=320$~Myr corresponds to the
distance of $x=$266~kpc. }\label{fig:evo}
\end{figure}

We solve equation~(\ref{eq:evo}) assuming that the whole relic is
steady. This means that electrons are constantly accelerated at the
shock and are reaccelerated in the turbulence. The CR electrons are
swept with the ICM and those at a distance $x$ from the shock were
located at the shock $x/V_{\rm d}$ ago. Figure~\ref{fig:evo} shows the
evolution of the CR electron spectrum on the comoving frame. At $t=0$,
the spectrum is given by equation~(\ref{eq:DSA}) and $s=2.6$. At $t>0$,
the electrons are reaccelerated in the turbulence and the number of
high-energy electrons ($\gamma\gtrsim 10^4$) increases.  Those electrons
are responsible for the observed synchrotron emission. The acceleration
time scale is given by $t_{\rm acc}\sim p^2/D_{pp}$, and it is $\sim
8$~Myr at $x=0$ and $p=p_0$. At $t\gtrsim 20$~Myr, the turbulence
significantly decays because $t>t_0=12$~Myr and then cooling becomes
effective. The cooling time is $t_{\rm cool}\sim p/(dp/dt)\sim
10$--100~Myr for $\gamma\sim 10^4$--$10^5$. Thus, the number of
high-energy electrons decreases.

\begin{figure}
 \begin{center}
  \includegraphics[width=8cm]{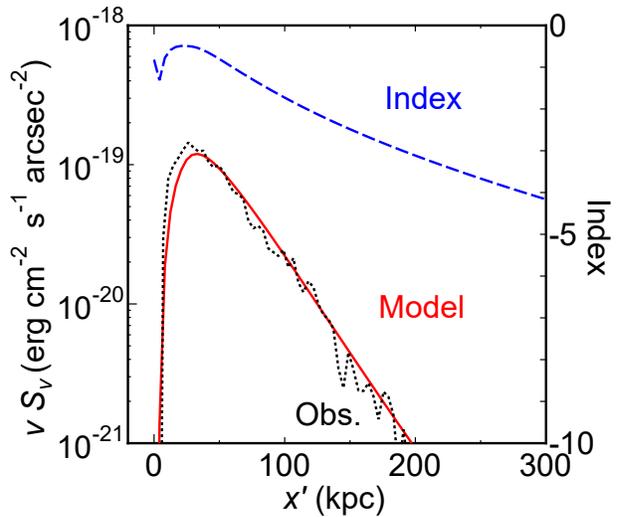} 
 \end{center}
\caption{Synchrotron surface brightness at 610~MHz as a function of the
projected distance from the shock. The solid curve is our prediction and
the dotted curve is the observation \citep{van10a}. The normalization of
the observed profile is adjusted to be consistent with the total flux
(0.2223~Jy) obtained by \citet{str16a}. The dashed curve is our
prediction for the spectral index between 153~MHz and
2272~MHz.}\label{fig:prof}
\end{figure}

In figure~\ref{fig:prof}, we show the surface brightness at 610~MHz as a
function of the projected distance from the shock ($x'$).  For that
purpose, we transform the planar shock we implicitly assumed into a
curved shock, and then project the radio emission along the line of
sight on the sky. Following previous studies, we assume that the shock
has a ribbon-like curved surface and it is viewed edge-on with a viewing
extension angle of $10^\circ$ (the total angle subtend is $20^\circ$;
\cite{van10a,kan15b}). The curvature radius is 1.5~Mpc.  As can be seen,
the model profile reproduces the observations well \citep{van10a}. At
$x=0$ and $p=p_0$, the spatial scale of the reacceleration is $x_{\rm
acc}=V_{\rm d} t_{\rm acc}\sim 7$~kpc, which is a factor of a few
smaller than the peak position of the surface brightness ($x'\sim
20$--$30$~kpc) reflecting the increasing $\eta(t)$
[equation~(\ref{eq:eta})].  We note that the slower rise of the radio
profile at $x'\gtrsim 0$ does not necessarily mean that our
reacceleration model is superior to DSA models. For the latter, the
radio emissivity generally has its maximum at the shock
($x=0$). However, if the shock is curved, the peak position is shifted
to $x'>0$ (see figure 4 in \cite{van10a}).  We also show the profile of
spectral index $\alpha$ between 153 and 2272~MHz in
figure~\ref{fig:prof}. At $x=0$, the index is $\alpha_{\rm DSA}=-0.82$,
which is the prediction based on the DSA for ${\cal M}=2.7$:
\begin{equation}
\label{eq:alpha}
 \alpha_{\rm DSA}=\frac{1}{2}-\frac{{\cal M}^2 + 1}{{\cal M}^2 - 1}\:
\end{equation}
(e.g. \cite{bla87a}). At $x'\gtrsim 0$, the index decreases a little
because the momentum spectrum at $\gamma\gtrsim 10^4$ is soft at the
beginning of reacceleration (see $t=5$~Mpc in figure~\ref{fig:evo}).
Then it starts increasing as the reacceleration proceeds. It would be
difficult to observe this dip of the index because of the low surface
brightness at $x'\sim 0$ (figure~\ref{fig:prof}). The index reaches the
maximum, $\alpha\sim -0.5$, at $x'\sim 20$~kpc, around which the surface
brightness is also maximized. The maximum index is comparable to
observation ($\alpha\sim -0.6$), which is inconsistent with the standard
DSA model because it predicts ${\cal M}\sim 4.6$ from $\alpha\sim -0.6$
\citep{van10a,str13a}.  Then, the index and surface brightness gradually
decrease owing to the decay of turbulence and cooling.

\begin{figure}
 \begin{center}
  \includegraphics[width=8cm]{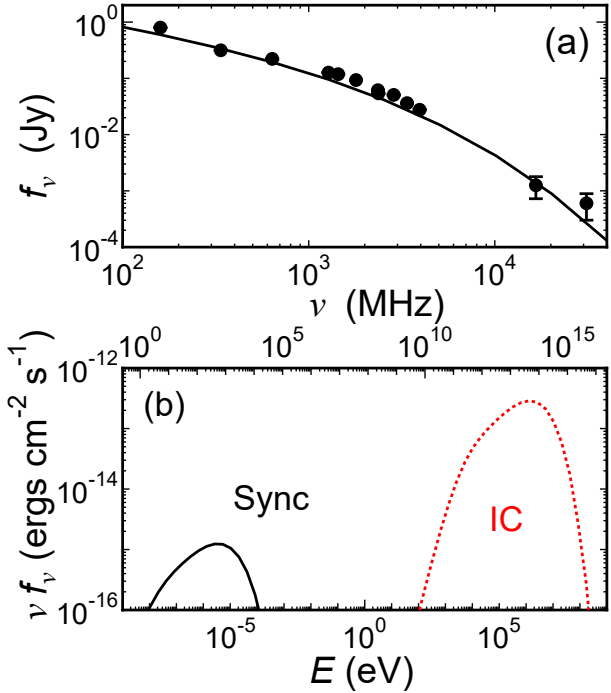} 
 \end{center}
\caption{Results for $B_{\rm
d}=0.30\:\mu\rm G$, $\eta_{\rm min}=1.6\times 10^{-7}$, and
$\chi_e=5.0\times 10^{-7}$. (a) Integrated radio spectrum (solid
line). Filled circles show the data from \citet{str16a}. (b) Broad band
spectra. The solid line shows synchrotron emission and the dotted line
shows inverse Compton scattering.}  \label{fig:spec}
\end{figure}

In figure~\ref{fig:spec}a, we show the radio spectrum of the whole relic
and compare it with observations by \citet{str16a}.  We assume that the
length of the relic is 1.7~Mpc and the viewing extension angle is
$10^\circ$.  Our model reproduces the curved spectrum very well, because
the momentum spectrum of the CR electrons is not a power law
(figure~\ref{fig:evo}).  The curved spectrum is a natural consequence of
second-order Fermi acceleration affected by cooling and a finite
acceleration time, which is of the order of $t_0$ in our case. In
figure~\ref{fig:spec}b, we show the broad-band spectra of the relic.
The hard X-ray flux from inverse Compton scattering is larger than the
synchrotron flux, because the magnetic field of $B_{\rm d}=0.30\:\mu\rm
G$ is smaller than $B_{\rm CMB}\approx 3.24 (1+z)^2\:\mu\rm G\approx
4.6\:\mu\rm G$, at which synchrotron cooling is comparable to the
cooling by inverse Compton scattering.\footnote{ Since the spectra of CR
electrons are not represented by a power law (figure~\ref{fig:evo}), the
relation between the synchrotron flux, the inverse Compton flux, and the
magnetic fields cannot be discussed using the usual formula based on a
power-law spectrum (e.g. equation (5.10) in \cite{sar86a}).} We analyze
the Suzaku data of the Sausage Cluster (see \cite{aka15a}) and find that
the upper limit of the non-thermal emission is $8.2\times 10^{-13}\rm\:
erg\: s^{-1}$ in the 0.3--10~keV band, which is consistent with our
prediction in figure~\ref{fig:spec}b.

\begin{figure}
 \begin{center}
  \includegraphics[width=8cm]{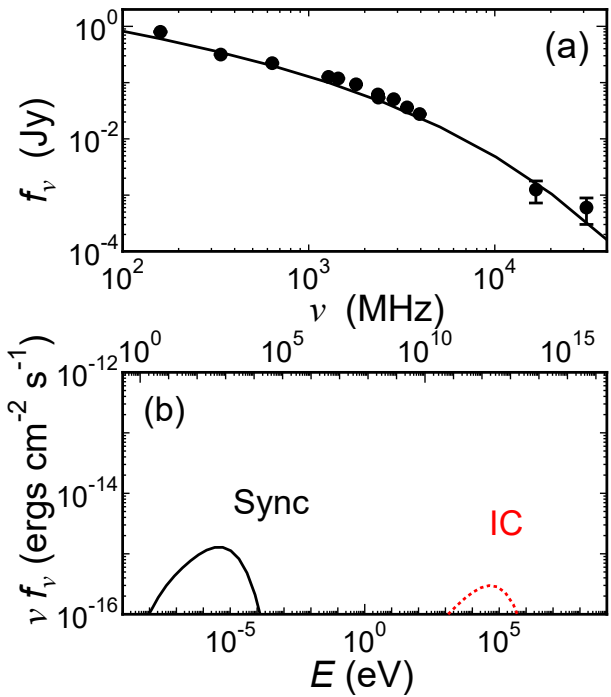} 
 \end{center}
\caption{As figure~\ref{fig:spec} but for $B_{\rm d}=9.5\:\mu\rm G$,
$\eta_{\rm min}=2.9\times 10^{-4}$, and $\chi_e=1.5\times 10^{-8}$.}
\label{fig:spec2}
\end{figure}

The width of the relic $\Delta x$ depends on the cooling time of CR
electrons. The typical frequency of synchrotron emission from an
electron with a Lorentz factor $\gamma$ in a magnetic field $B$ is
$\nu_c\propto\gamma^2 B$. The cooling time of the electron is $t_{\rm
cool}\propto \gamma^{-1}(B^2 + B_{\rm CMB}^2)^{-1} \propto
B^{1/2}\nu_c^{-1/2}(B^2 + B_{\rm CMB}^2)^{-1}$
(e.g. \cite{ryb79}). Thus, two values of $B$ give the same $t_{\rm
cool}$ for a given $\nu_c$. This means that for a given observation
frequency $\nu_{\rm obs}\sim\nu_c$, there are two values of $B$ that
give the same $\Delta x$.\footnote{Note that the width is not simply
given by $\Delta x= V_{\rm d}t_{\rm cool}$, which is often assumed when
only cooling is important (e.g. \cite{van10a}), because it also depends
on the reacceleration of CRs at $x>0$. Thus, the magnetic field that
gives a certain $\Delta x$ depends on whether reacceleration is
considered or not.}  For the Sausage Cluster, the value we adopted in
figure~\ref{fig:spec} ($B_{\rm d}=0.30\:\mu\rm G$) is the smaller
one. Figure~\ref{fig:spec2} is the other solution with the larger
magnetic field ($B_{\rm d}=9.5\:\mu\rm G$). The other parameters are the
same as those for figure~\ref{fig:spec} except for $\eta_{\rm
min}=2.9\times 10^{-4}$, and $\chi_e=1.5\times 10^{-8}$.
Figure~\ref{fig:spec2}a is almost the same as
figure~\ref{fig:spec}a. Moreover, we have confirmed that the profiles of
surface brightness and spectral index for this large $B_{\rm d}$ model
are almost identical to those in figure~\ref{fig:prof}. On the other
hand, the hard X-ray flux in figure~\ref{fig:spec2}b is much smaller
than that in figure~\ref{fig:spec}b, which can be used to discriminate
the high and low $B_{\rm d}$ models in future observations. Note
that in this large $B_{\rm d}$ model, the spatial diffusion time ($\sim
x^2/D_{xx}$) of the CRs is comparable to the advection time ($\sim
x/V_{\rm d}$) around the peak of the surface brightness ($x\sim
20$--30~kpc). Thus, the best fit parameters we obtained above would be
somewhat changed if we include the effect of spatial diffusion.

\section{Summary and Discussion}

In this study, we have shown that turbulent reacceleration of electrons
can explain the observed surface brightness profile and the hard, curved
radio spectrum of the northern radio relic in the Sausage Cluster. This
model also reproduces the properties of the relic in the Toothbrush
Cluster (1RXS~J0603.3$+$4214l; \cite{fuj15b}). So far, curved radio
spectra have been observed for the bright relics in three clusters
(Sausage, Toothbrush, and A~2256; \cite{tra15a}).  This may indicate
that the turbulent reacceleration is a fairly common mechanism to create
radio relics in clusters. The spatial scale of turbulence must be
smaller than the width of the relic and the mean free path of CR
electrons must be much shorter than the Coulomb mean free path of
thermal particles.

In our model, we assumed that the efficiency of turbulent reacceleration
is the highest at the shock front [equation~(\ref{eq:eta})].  However,
if it takes some time for eddies to grow up, the turbulent
reacceleration starts to develop well behind the shock and the
synchrotron emission would clearly be detached from the shock. If this
is observed in some relics, it would support our model. Moreover, X-ray
observations with a high spectral resolution (e.g., Hitomi, Athena) may
reveal turbulence behind shocks.

Some key issues remain unsolved. First, the generation of turbulence
needs to be addressed. If a shock travels through inhomogeneous gas,
turbulence may be generated \citep{sam94a,ino09a}. In fact, some recent
X-ray observations suggest that the ICM in the outskirts of clusters is
clumpy \citep{sim11a}. Using numerical simulations, \citet{van11b}
concluded that the amplitude of density fluctuation must be $\lesssim
30$ percent. However, the wavelengths of the fluctuation they studied
are rather large ($\geq 75$~kpc), and the evolution of turbulence when
the wavelength is smaller (say 10~kpc) is not certain. Alternatively, if
a shock passes through a turbulent region, turbulence could be amplified
downstream of the shock \citep{iap12a}.  Moreover, some plasma
instabilities may be developing and generating turbulence
downstream. Second, relics are often polarized \citep{fer12a}, which may
be incompatible with the turbulent reacceleration model. For the Sausage
Cluster, the relic is strongly polarized at the 50 to 60\% level
\citep{van10a}. However, compressional amplification of the magnetic
field may account for a large polarization of the radio emission
\citep{iap12a}. Note that in our model the mean free path associated
with the turbulent reacceleration ($l_{\rm mfp}$) is much larger than
the gyro radius of the CR electrons:
\begin{equation}
 r_{\rm g} = \frac{p c}{e B_{\rm d}} \approx 5.7\times 10^{13} 
\left(\frac{p}{10^4\: m_e c}\right)
\left(\frac{B_{\rm d}}{0.30\:\mu\rm G}\right)^{-1}\:\rm cm\:,
\end{equation}
where $e$ is the elementary charge. In general, the mean free path is
represented as $l_{\rm mfp}\sim r_{\rm g}(B/\delta B)^2$, where $r_{\rm
g}$ is the gyro radius of a particle, $B$ is the background magnetic
field, and $\delta B$ is the fluctuation of the field
(e.g. \cite{lon94a}). Thus, $l_{\rm mfp}\gg r_{\rm g}$ means $\delta
B/B\ll 1$, which may indicate that the turbulence on the micro-scales is
not strong and magnetic fields may be organized enough. If the model
with the larger $B_{\rm d}$ and $\eta_{\rm min}$ is realized
(figure~\ref{fig:spec2}), the fluctuation $\delta B/B$ is even smaller.

\begin{ack}
We thank the anonymous referee for a thorough review and constructive
suggestions.  We are grateful to R.~J. van Weeren for providing us
observational data. This work was supported by KAKENHI No.~15K05080
(Y.~F.). H.~A. is supported by a Grant-in-Aid for Japan Society for the
Promotion of Science (JSPS) Fellows (26-606).
\end{ack}



\begin{thebibliography}{}

\bibitem[Akamatsu et al.(2012a)]{aka12a} Akamatsu, H., de Plaa, 
J., Kaastra, J., et al.\ 2012, \pasj, 64, 49

\bibitem[Akamatsu 
\& Kawahara(2013)]{aka13a} Akamatsu, H., \& Kawahara, H.\ 2013, \pasj,
		65, 16

\bibitem[Akamatsu et al.(2012b)]{aka12b} Akamatsu, H., 
Takizawa, M., Nakazawa, K., et al.\ 2012, \pasj, 64, 67

\bibitem[Akamatsu et 
al.(2015)]{aka15a} Akamatsu, H., van Weeren, R.~J., Ogrean, G.~A., et
		al.\ 2015, \aap, 582, A87

\bibitem[Basu et al.(2015)]{bas15a} Basu, K., Vazza, F., 
Erler, J., \& Sommer, M.\ 2015, arXiv:1511.03245 

\bibitem[Blandford 
\& Eichler(1987)]{bla87a} Blandford, R., \& Eichler, D.\
		1987, \physrep, 154, 1 

\bibitem[Bourdin et al.(2013)]{bou13a} Bourdin, H., Mazzotta, 
P., Markevitch, M., Giacintucci, S., \& Brunetti, G.\ 2013, \apj, 764,
		82

\bibitem[Brunetti et al.(2004)]{bru04a} Brunetti, G., Blasi, 
P., Cassano, R., \& Gabici, S.\ 2004, \mnras, 350, 1174 

\bibitem[Brunetti 
\& Jones(2014)]{bru14a} Brunetti, G., \& Jones, T.~W.\ 2014,
		International Journal of Modern Physics D, 23, 1430007


\bibitem[Brunetti 
\& Lazarian(2007)]{bru07a} Brunetti, G., \& Lazarian, A.\ 2007, \mnras,
		378, 245

\bibitem[Brunetti et al.(2001)]{bru01a} Brunetti, G., Setti, 
G., Feretti, L., \& Giovannini, G.\ 2001, \mnras, 320, 365 

\bibitem[Donnert et al.(2016)]{don16} Donnert, J.~M.~F.,
		Stroe, A., Brunetti, G., Hoang, D., \& Roettgering, H.\
		2016, arXiv:1603.06570

\bibitem[Feretti et 
al.(2012)]{fer12a} Feretti, L., Giovannini, G., Govoni, F., \& Murgia,
		M.\ 2012, \aapr, 20, 54

\bibitem[Finoguenov et al.(2010)]{fin10} Finoguenov, A., 
Sarazin, C.~L., Nakazawa, K., Wik, D.~R., 
\& Clarke, T.~E.\ 2010, \apj, 715, 1143 

\bibitem[Fujita 
\& Sarazin(2001)]{fuj01b} Fujita, Y., \& Sarazin, C.~L.\ 2001, \apj,
		563, 660

\bibitem[Fujita et al.(2003)]{fuj03a} Fujita, Y., Takizawa, 
M., \& Sarazin, C.~L.\ 2003, \apj, 584, 190 

\bibitem[Fujita et al.(2015)]{fuj15b} Fujita, Y., Takizawa, 
M., Yamazaki, R., Akamatsu, H., \& Ohno, H.\ 2015, \apj, 815, 116 

\bibitem[Gould(1975)]{gou75a} Gould, R.~J.\ 1975, \apj, 196, 
689 

\bibitem[Inoue et al.(2009)]{ino09a} Inoue, T., Yamazaki, R., 
\& Inutsuka, S.\ 2009, \apj, 695, 825 

\bibitem[Inoue et al.(2010)]{ino10a} Inoue, T., Yamazaki, R., 
\& Inutsuka, S.-i.\ 2010, \apjl, 723, L108 

\bibitem[Iapichino 
\& Br{\"u}ggen(2012)]{iap12a} Iapichino, L., \& Br{\"u}ggen, M.\ 2012,
		\mnras, 423, 2781

\bibitem[Isenberg(1987)]{ise87a} Isenberg, P.~A.\ 1987, \jgr, 
92, 1067 

\bibitem[Itahana et al.(2015)]{ita15a} Itahana, M., Takizawa, 
M., Akamatsu, H., et al.\ 2015, \pasj, 67, 113

\bibitem[Kang et al.(2012)]{kan12a} Kang, H., Ryu, D., 
\& Jones, T.~W.\ 2012, \apj, 756, 97 

\bibitem[Kang 
\& Ryu(2015)]{kan15b} Kang, H., \& Ryu, D.\ 2015, \apj, 809, 186 

\bibitem[Longair(1994)]{lon94a}
Longair M. S., 1994, High Energy Astrophysics, Vol. 2, 2nd edn. Cambridge
Univ. Press, Cambridge. Section 20.4

\bibitem[Macario et al.(2011)]{mac11a} Macario, G., 
Markevitch, M., Giacintucci, S., et al.\ 2011, \apj, 728, 82 

\bibitem[Mertsch \& Sarkar(2011)]{mer11a} Mertsch, P., \&
Sarkar, S.\ 2011, Physical Review Letters, 107, 091101

\bibitem[Ogrean 
\& Br{\"u}ggen(2013)]{ogr13b} Ogrean, G.~A., \& Br{\"u}ggen, M.\ 2013,
		\mnras, 433, 1701

\bibitem[Ogrean et al.(2013)]{ogr13a} Ogrean, G.~A., 
Br{\"u}ggen, M., van Weeren, R.~J., et al.\ 2013, \mnras, 433, 812 

\bibitem[Ohno et al.(2002)]{ohn02a} Ohno, H., Takizawa, M., 
\& Shibata, S.\ 2002, \apj, 577, 658 

\bibitem[Petrosian(2001)]{pet01b} Petrosian, V.\ 2001, \apj, 
557, 560 

\bibitem[Roettiger et al.(1999)]{roe99a} Roettiger, K., Burns, 
J.~O., \& Stone, J.~M.\ 1999, \apj, 518, 603 

\bibitem[Rybicki
\& Lightman(1979)]{ryb79}
Rybicki G.~B., \& Lightman, A.~P. 1979, Radiative Processes in
	       Astrophysics
(New York: Wiley)

\bibitem[Samtaney \& Zabusky(1994)]{sam94a}
Samtaney R., \& Zabusky N. J.,\ 1994, J. Fluid Mech., 269, 45

\bibitem[Sarazin(1986)]{sar86a} Sarazin, C.~L.\ 1986, Reviews 
of Modern Physics, 58, 1 

\bibitem[Sarazin(1999)]{sar99a} Sarazin, C.~L.\ 1999, \apj, 
520, 529 

\bibitem[Sasaki et al.(2015)]{sas15a} Sasaki, K., Asano, K., 
\& Terasawa, T.\ 2015, \apj, 814, 93 

\bibitem[Schlickeiser(1989)]{sch89a} Schlickeiser, R.\ 1989, 
\apj, 336, 243 

\bibitem[Schlickeiser et 
al.(1987)]{sch87a} Schlickeiser, R., Sievers, A., \& Thiemann, H.\ 1987,
		\aap, 182, 21

\bibitem[Simionescu et al.(2011)]{sim11a} Simionescu, A., 
Allen, S.~W., Mantz, A., et al.\ 2011, Science, 331, 1576 

\bibitem[Stroe et al.(2016)]{str16a} Stroe, A., Shimwell, T., 
Rumsey, C., et al.\ 2016, \mnras, 455, 2402 

\bibitem[Stroe et 
al.(2013)]{str13a} Stroe, A., van Weeren, R.~J., Intema, H.~T., et al.\
		2013, \aap, 555, A110

\bibitem[Sturner et al.(1997)]{stu97a} Sturner, S.~J., Skibo, 
J.~G., Dermer, C.~D., \& Mattox, J.~R.\ 1997, \apj, 490, 619 

\bibitem[Trasatti et 
al.(2015)]{tra15a} Trasatti, M., Akamatsu, H., Lovisari, L., et al.\
		2015, \aap, 575, A45

\bibitem[van Weeren et al.(2016)]{van16a} van Weeren, R.~J., 
Brunetti, G., Br{\"u}ggen, M., et al.\ 2016, arXiv:1601.06029 

\bibitem[van Weeren et al.(2011)]{van11b} van Weeren, R.~J., 
Br{\"u}ggen, M., R{\"o}ttgering, H.~J.~A., 
\& Hoeft, M.\ 2011, \mnras, 418, 230 

\bibitem[van Weeren et al.(2010)]{van10a} van Weeren, R.~J., 
R{\"o}ttgering, H.~J.~A., Br{\"u}ggen, M., 
\& Hoeft, M.\ 2010, Science, 330, 347 

\bibitem[van Weeren et 
al.(2012)]{van12b} van Weeren, R.~J., R{\"o}ttgering,
	  H.~J.~A., Intema, H.~T., et al.\ 2012, \aap, 546, A124 

\bibitem[Yamazaki 
\& Loeb(2015)]{yam15a} Yamazaki, R., \& Loeb, A.\ 2015, \mnras, 453,
		1990

\bibitem[Yang 
\& Zhang(2009)]{yan09} Yang, X.~C., \& Zhang, L.\ 2009, \aap, 496, 751 

\bibitem[Zhang(2015)]{zha15a} Zhang, M.\ 2015, \apj, 812, 148 

\bibitem[ZuHone et al.(2013)]{zuh13a} ZuHone, J.~A., 
Markevitch, M., Brunetti, G., \& Giacintucci, S.\ 2013, \apj, 762, 78 


\end{thebibliography}
\end{document}